\documentclass[a4paper]{mem}
\usepackage{natbib}
\usepackage{graphicx}
\usepackage[a4paper]{hyperref}
\idline{73}{23}
\def\simgr{\,\hbox{\hbox{$ > $}\kern -0.8em \lower 1.0ex\hbox{$\sim$}}\,}
\def\simle{\,\hbox{\hbox{$ < $}\kern -0.8em \lower 1.0ex\hbox{$\sim$}}\,}

\begin{document}
   \title{Pulsations induced by deuterium-burning in young 
   brown dwarfs }

   \author{Francesco Palla \inst{1}
   \and Isabelle Baraffe \inst{2} }

   \offprints{F. Palla}

   \institute{INAF-Osservatorio Astrofisico di Arcetri, L.go E. Fermi 5,
   50125 Firenze, Italy \email{palla@arcetri.astro.it} \\ 
              \and  CRAL-CNRS, Ecole Normale Sup\'erieure, 69364 Lyon Cedex 
   07, France \\ \email{ibaraffe@ens-lyon.fr}
             }

   \abstract{Very low-mass stars (VLMSs) and
   brown dwarfs (BDs) can undergo pulsational instability excited by central
   deuterium burning during the initial phases of their evolution. We
   present the results of evolutionary and nonadiabatic linear stability 
   models that show the presence of unstable fundamental modes. The 
   pulsation periods vary bewteen $\sim$5 hr for a 0.1 M$_\odot$ star and
   $\sim$1 hr for a 0.02 M$_\odot$ brown dwarf. 
   The results are rather insensitive to
   variations in the input physics of the models. We show the location of
   the instability strip in the HR and c-m diagrams and discuss the 
   observational searches for young pulsators in nearby star forming regions.
   
   }
   \authorrunning{F. Palla \& I. Baraffe}
   \titlerunning{Pulsations in young BDs}
   \maketitle
%

\section{Introduction}

Evidence for photometric variability of young VLMSs and BDs 
in star forming regions has been obtained by several groups in the last
few years (Bailer Jones \& Mundt 2001, Joergens et al. 2003, Zapatero
Osorio et al. 2003; see also Eisl\"offel in this volume). The amplitudes of
the observed variations range from tens of mmag in the optical to
$\sim$0.05--0.2 mag in the near-IR where these cool objects emit most of
their energy.  In several cases,  periodic variability has been reported with
tentative periods in the range from half an hour to several hours. Very
interesting results have been found recently in the $\epsilon$ and $\sigma$ 
Orionis clusters with several objects showing periods in this range
(e.g. Caballero et al. 2004, Scholz \& Eisl\"offel 2004). So
far, the observed variability has been interpreted in terms of rotation
periods, presence of photospheric cool or hot spots, interaction with
accretion disks, and/or atmospheric events.  However, for the objects with
the shortest periods ($\simle$1~hr), the inferred rotational velocities would
exceed $\sim$100 km~s$^{-1}$ the breakup speed where gravitational and
centrifugal forces balance, thus making the rotation interpretation less
likely.

In this contribution, we present the initial results of a study of the
stability of young VLMSs and BDs during the initial phases of their
contraction. In particular, we find that once the central temperatures allow
the ignition of deuterium (D-)burning (T$\sim 10^6$~K), the whole
interior can become {\it pulsationally unstable} due to the high sensitivity
on temperature of the energy generation rate. Such instability, called
$\epsilon$-mechanism, was originally suggested by Gabriel (1964) to occur in
fully convective, low-mass stars on the main sequence. Later on, Toma (1972)
showed that PMS stars in the range 0.2--2.0 M$_\odot$ could also become
pulsationally unstable during the D-burning phase and suggested a possible
relation of this phenomenon with the observed variability of T Tauri stars.
In both cases, and for different reasons, the original suggestion has not met
with success and to date there is no observational evidence for the existence of
the $\epsilon$-mechanism in any class of objects. As we will see below, VLMSs
and BDs in the earliest evolutionary stages have the appropriate physical
conditions to undergo D-induced instabilities on time scales from few hours
to about an hour, similar to those observed in some objects in nearby star
forming regions (SFRs).

\section {The early evolution of VLMSs and BDs}
These objects are expected to begin the PMS phase with convective interiors
and with the full amount of interstellar
 deuterium since during the preceding phase of protostellar accretion their
 centers are too cold to start any nuclear reaction (Stahler 1988). In fact,
 the ignition temperature of $\sim 10^6$~K can be reached only in protostars
 more massive than $\sim$0.2 $M_\odot$, depending on the accretion history.
 Thus, lower mass objects need to contract somewhat in the PMS phase before
 deuterium can start burning and being depleted. Once the critical
 tempearture is achieved, the D-burning phase occurs on a time scale that
 varies between $t_{\rm burn}\sim$2 Myr for a 0.1~M$_\odot$ star and $t_{\rm
 burn}\sim$20 Myr for a 0.02~M$_\odot$ brown dwarf (Baraffe et al. 1998).
 Since the energy generation rate, $\epsilon$, scales approximately with the
 12-th power of temperature, the rate of combustion is slower for the lowest
 mass objects.  The same temperature sensitivity of $\epsilon$ is at the root
 of the instability induced by any temperature variation: since $\delta
 \epsilon/\epsilon \sim 12 \,\delta T/T$, a small T-perturbation induces a
 variation of $\epsilon$ which is an order of magnitude bigger.  In terms of
 pulsation analysis, the time scale for the growth of the instability,
 $\tau_d$, is inversely proportional to $\delta \epsilon$ and should be
 shorter than the D-burning time, $t_{\rm burn}$, for the mechanism to
 operate.  Therefore, in order to test its viability, it is important to
 follow numerically the growth of the instability in time since the onset of
 the D-burning phase.

\section {Evolution and stability of VLMSs and BDs}
Evolutionary models of objects with mass between 0.1 and 0.02 M$_\odot$ have
been computed using the Lyon code that provides a careful treatement of both
the inner structure and the external layers (atmosphere).  The initial D-mass
fraction is set equal to $2\times 10^{-5}$ (e.g. Linsky 1998) and the
atmosphere is assumed to be dust free since BDs are relatively hot during the
initial evolution (T$_{\rm eff}\simgr$2300~K).

For the nonadiabatic, linear stability analysis we have used the models of
Baraffe et al. (2001) that search for the presence of unstable radial eigen
modes, characterizing oscillations around the hydrostatic equilibrium
configuration. If the perturbations have time to grow, they could reach large
amplitudes and result in  periodic phases of expansion and contraction, with
a pulsation period  related to the dynamical time scale of the object
$\tau_{\rm dyn} \sim (G \bar \rho)^{-1/2}$.

\section{Results}

The main results of the stability analysis (Palla \& Baraffe 2004) are the
following:  (1) D-burning does induce the expected pulsations; (2) the period
of the fundamental mode of the pulsation is short and varies between
$\sim$4~hr for a 0.1 M$_\odot$ star and $\sim$1~hr for a 0.02 M$_\odot$ BD;
(3) the growth time is short compared to the evolutionary time and decreases
substantially for smaller masses. The models find a ratio $\tau_d/t_{\rm
burn}\simle$10 in the BD regime, satisfying the requirement for the growth of
the instability.  Therefore, the mechanism can operate effectively before all
the available deuterium is consummed.

\begin{figure}
   \centering
   \includegraphics[width=7cm,height=8cm]{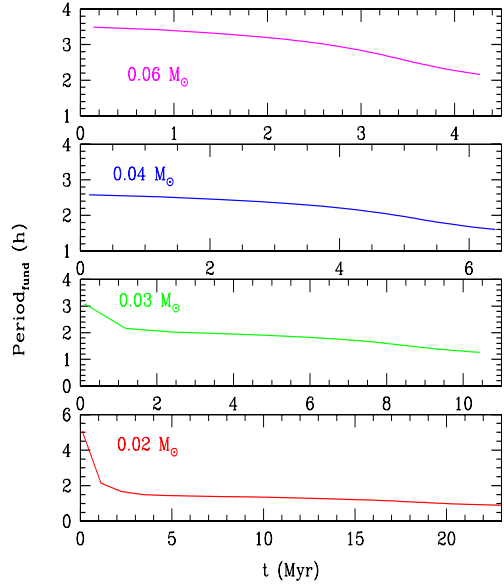}
   \caption{Variation of the period of the fundamental mode with time for
   BD models in the mass interval 0.06 to 0.02 M$_\odot$. Note the different
   tome scale in the various panels.}
   \label{fig1}
   \end{figure}

\begin{figure}
   \centering
   \includegraphics[width=7cm,height=8cm]{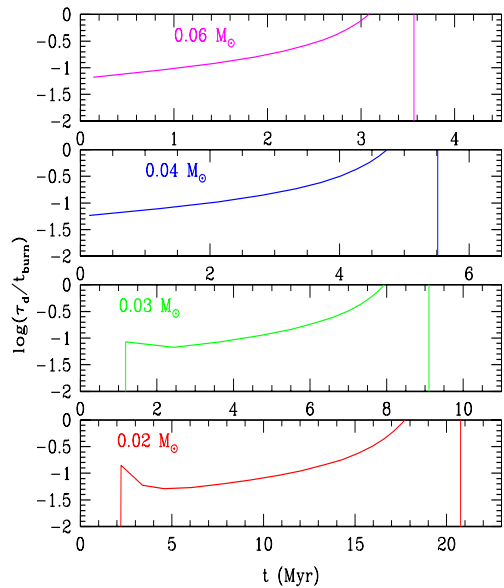}
   \caption{Time variation of the ratio of the D-burning and growth times
   for the same models of Fig.~\ref{fig1}.}
   \label{fig2}
   \end{figure}

Fig.~\ref{fig1} displays the time variation of the period of the fundamental
mode for four BD models. In all cases, the pulsation period remains constant
during the main D-burning phase and slowly decreases with time in the final
stage of complete depletion.  The comparison between the growth time of the
oscillation and the D-burning time for the same models is displayed in
Fig.~\ref{fig2}.  Note how the growth time remains well below $t_{\rm burn}$,
thus guaranteeing that the oscillation can grow to significant amplitudes.
Unfortunately, the results of the linear analysis cannot be used to calculate
the expected amplitudes of the pulsations. For this, a nonlinear study should
be carried out, but these (very complicated) models are still not available.
However, considering that the perturbation is produced in the deep interior,
we expect small amplitudes in all cases (below $\sim$0.1 mag).

\begin{figure}
   \centering
   \includegraphics[width=6.5cm,height=7cm]{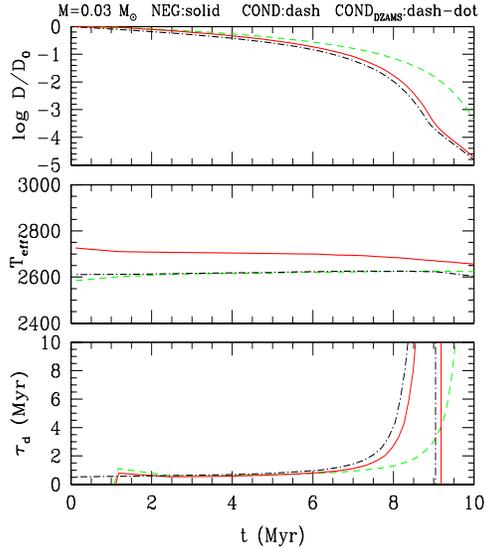}
   \caption{Effects of variations in the input physics in the 0.03 M$_\odot$
   case. Atmospheric models:
   NEG (solid) vs. COND (dash). The 
   dash-dotted line is for a COND model with initial central temperature 
   close to the D-ignition value.}
   \label{fig3}
   \end{figure}

The results on the instability are robust against uncertaintites in 
the stellar input physics, such as atmospheric models (NextGen vs.
COND models which use different molecular linelists), nuclear reaction rates
(Caughlan \& Flower vs. NACRE), and initial conditions (above or at the
D-ignition).  Changes in these quantities may affect the duration of
the D-burning phase or the growth time of the instability, but the
pulsational properties remain the same.  On the other hand, the treatment of
convection is a delicate issue, since convection is expected to damp the
acoustic waves.  However, since the convective time scale ($\gg$yr) is much
longer than the pulsation period ($\sim$hr), the standard assumption that
convection is frozen with the pulsation appears reasonable.  To appreciate
the (in)sensitivity of the results on the assumed atmospheric model and
initial central temperature (below D-ignition), Fig.~\ref{fig3} shows the
variation of various quantities for the 0.03 M$_\odot$ case. Both the
duration of the D-burning phase and the magnitude of the pulsation growth
time remain basically unchanged.

\section {The D-instability strip in the HR and CM diagrams}

\begin{figure}
   \centering
   \includegraphics[width=6.5cm]{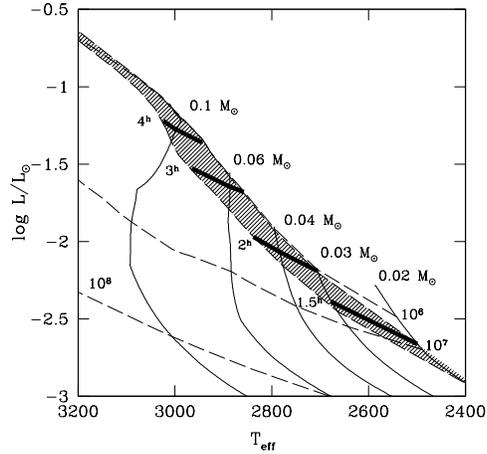}
   \caption{Location of the D-instability strip in the HR diagram. 
   Evolutionary tracks for different masses are as labeled. Selected 
   isochrones are for $10^6$, $10^7$, and $10^8$~yr. The heavy solid lines
   represent the curves of constant period for 4, 3, 2 and 1.5 hr.}
   \label{fig4}
   \end{figure}

\begin{figure}
   \centering
   \includegraphics[width=6.5cm]{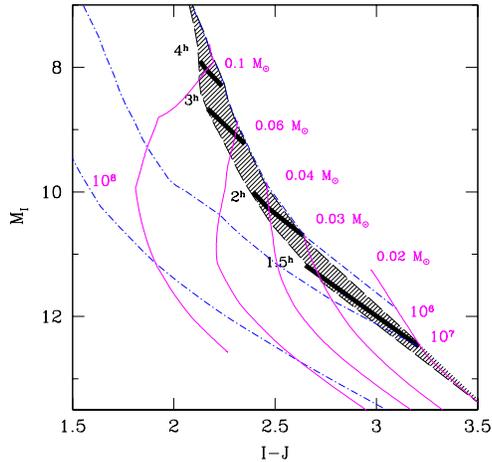}
   \caption{The D-instability strip in the ($I,I-J$) diagram. Heavy solid 
   lines have the same meaning as in Fig.~\ref{fig4}.}
   \label{fig5}
   \end{figure}

 The location of the D-instability strip in the HR diagram is shown in 
 Fig.~\ref{fig4}, along with tracks of selected
 masses and isochrones. The curves of constant period of the fundamental
 mode cut the strip almost
 horizontally at nearly constant luminosity and the shortest periods are found
 for the least massive BDs. 
 Since the evolutionary models also provide the flux emitted by the BDs at
 different wavelengths, we show the corresponding color-magnitude diagram in 
 Fig.~\ref{fig5} for the $I-$ and $J-$bands. Both plots can be conveniently
 used for comparison with observations.

\section {The observational search for young pulsators}

As a result of deep and extended surveys both in the optical and near-IR,
young brown dwarfs have been discovered in large numbers in nearby clusters
and associations (e.g., Mart\'{\i}n et al. 2004, L\'opez Mart\'{\i} et al.
2004, Slesnick et al. 2004) with the conclusion that there are almost as many
BDs as regular stars (Chabrier 2004).  Since age spreads in SFRs
are similar to the duration of the D-burning phase in VLMSs and BDs
(between a few Myr and $\simle$20 Myr), there is ample choice of potential
candidates for the discovery of pulsations, despite the relatively narrow
extent of the instability strip (see Fig.~\ref{fig4} and Fig.~\ref{fig5}.)

\begin{figure}
   \centering
   \includegraphics[width=6.5cm]{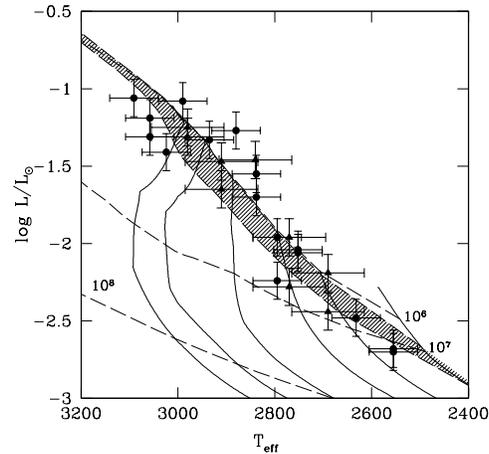}
   \caption{Distribution of known VLMSs and BDs in Taurus (circles) and Cha~I
   (triangles). Data points for Taurus are from (Brice\~no et al. (2002), 
   Luhman et al. (2003), Luhman (2004), and for Cha~I from Comer\'on et al. 
   (2000).}
   \label{fig6}
   \end{figure}

As an example, Fig.~\ref{fig6} displays the location in the HR diagram of 25
spectroscopically confirmed VLMSs and BDs drawn from larger surveys in Taurus
and Chamaeleon I (Brice\~no et al. 2002, Comer\'on et al. 2000, Luhman et
al.  2003, Luhman 2004).  The rather large error bars indicate the current
uncertainties in the conversion from spectral types to effective temperatures
but the overall distribution closely
matches the predicted position of the instability strip down to the lowest
masses with the shortest pulsational periods.  Interestingly, the sample of
BDs in Cha I does not show sign of residual accretion indicating that the
interaction with the circumstellar environment should not affect their
surface properties (Natta et al. 2004). This is an important advantage for
searches of intrinsic periodic variability.

The case for the Orion Nebula Cluster is also promising, particularly after
the spectroscopic confirmation of a rather large sample of VLMSs and BDs
(Slesnick et al. 2004). However, not all SFRs have the suitable population of
such objects. For example, the $\rho$~Ophiuchi BD sample appears too young
(Natta et al. 2004) with all BDs lying well above the D-instability strip.
Thus, care must be taken in the selection of the candidates for pulsation.

\begin{figure}
   \centering
   \includegraphics[width=6.5cm]{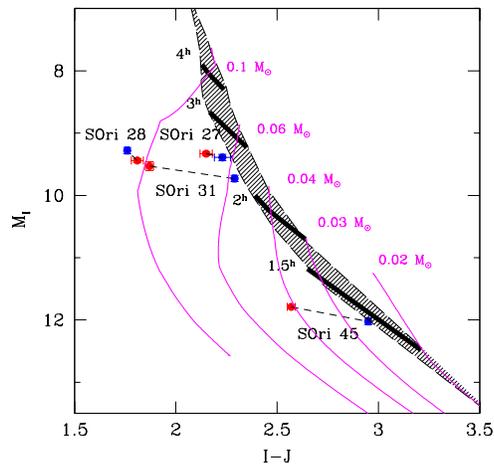}
   \caption{Position in the ($I,I-J$) diagram of the $\sigma$~Ori members with
   known periods.}
   \label{fig7}
   \end{figure}

In addition to global stellar properties, it is important to consider the
information already available on BD variability in young SFRs. The case for
$\sigma$~Orionis is particularly interesting for the relatively young age
($\sim$3 Myr) and the rich harvest of substellar objects.  Several multi-epoch
photometric studies of spectroscopically confirmed BDs have shown that
variability occurs in about 50\% of the sources on a variety of time scales
(from few hours to days and years, e.g.  Caballero et al. 2004). In several
cases, short-term periodic variations have also been found (or claimed). This
is the case for S~Ori 27 (period: 2.8$\pm$0.4 hr), S~Ori 28 (3.3$\pm$0.6 hr),
S~Ori 31 (1.8$\pm$0.2 hr), and S~Ori 45 (0.5$\pm$0.2 hr) (B\'ejar et al.
2001, 2004; Bailer-Jones \& Mundt 2001; Zapatero-Osorio et al. 2003;
Caballero et al.  2004).  The location of the four sources in the CM diagram
is shown in Fig.~\ref{fig7}.  The two data points for each source refer to
measurements from different epoch and underline the large uncertainty still
existing in the colors.  With the exception of S~Ori~28, all three BDs lie
very close to the instability strip and their estimated periods compare well
with the locus of the isoperiods, considering the errors on the period
determination.
In spite of the encouraging results, we would like to stress that owing to the
limited time coverage and 1-day alias problem that plague all variability
studies from ground-based observations, these periods are still to be
considered tentative and long-term, multi-band, multi-site monitoring in
the future are necessary to confirm their presence. On the other hand, 
more reliable measurements of the BD colors are also essential for a better
comparison with the predicted periods. In turn, these high quality data
can be used to place stronger constraints on the input physics of both the
evolutionary and stability models.

\section{Conclusions}
Perturbations excited in the center of D-burning VLMSs and BDs have time to
grow and persist during this ephemeral, but important nuclear interlude.
More theoretical work is required to confirm and extend the initial results
obtained here. In particular, it is important to develop nonlinear models
for the estimate of the pulsation amplitude and fully hydrodynamical
calculations to produce syntheic light curves to compare with observations.
In general, the reliable identification of the pulsation signature will not 
be an easy task, considering the other sources of periodic varabiltiy at
play in such objects. Notwithstanding these difficulties, the pulsational
instability induced by D-burning offers a new interpretation that relies on
fundamental (sub)stellar properties. As such, it has the potential
of providing direct information on the otherwise inaccesible internal
structure of VLMSs and BDs.  More fundamentally, it could provide the first
evidence for the existence of the $\epsilon$-mechanism in any astronomical
object.

\begin{acknowledgements}
We thank Antonella Natta and Leonardo Testi for the organization of an
exciting and interactive meeting. We acknowledge useful conversations on BDs
with J.  Eisl\"offel, D. Barrado y Navascu\'es and E. Masciadri.
\end{acknowledgements}

\bibliographystyle{aa}

\end{document}